\documentclass[prb,twocolumn,showpacs,amsmath,amssymb]{revtex4}
\usepackage{graphicx}% Include figure files
\usepackage{dcolumn}% Align table columns on decimal point
\usepackage{bm}% bold math

\begin{document}

%\preprint{APS/123-QED}

\title{Stationary phase-kink states and dynamical phase transitions controlled by\\
surface impedance in THz wave emission from intrinsic Josephson junctions}% Force line breaks with \\

\author{Yoshihiko Nonomura} \email{nonomura.yoshihiko@nims.go.jp}
\affiliation{%
Computational Materials Science Center, National Institute for Materials Science, 
Tsukuba, Ibaraki 305-0047, Japan}

\date{\today}% It is always \today, today,

\begin{abstract}
As possible states to characterize THz wave emission from intrinsic Josephson 
junctions without external fields, the McCumber-like state and $\pi$-phase-kink 
state have been proposed. In the present article it is numerically shown that both 
states are stationary according to the bias current $J$ and surface impedance $Z$. 
The McCumber-like state is stable for low $J$ and small $Z$. For higher $J$, the 
$\pi$-phase-kink state accompanied with symmetry breaking along the $c$ axis is 
stable even for $Z=1$, though strong emission in the vicinity of cavity resonance 
points only takes place for larger $Z$. Different emission behaviors for $Z=1$ and 
$10$ are precisely compared. The dynamical phase diagram for $1 \le Z \le 10$ 
and the optimal value of $Z$ for the strongest emission are also evaluated.
\end{abstract}
\maketitle

{\it Introduction.}
Emission from intrinsic Josephson junctions (IJJs) such as 
Bi$_{2}$Sr$_{2}$CaCu$_{2}$O$_{8}$ (BSCCO) has been intensively studied 
as a candidate of stable source of continuous terahertz electromagnetic wave. 
Although emission controlled by flow of Josephson vortices \cite{Koyama} 
had been investigated numerically, \cite{Machida,Tachiki,Lin1} 
experimental realization of such emission \cite{THz-exp} had been quite 
difficult. Recently more evident emission from IJJs without external fields 
was reported experimentally, \cite{Ozyuzer} and two types of states 
were proposed theoretically in order to explain the emission. 
One is the McCumber-like state (Ohmic and translational 
invariant along the $c$ axis), \cite{Matsumoto} and another 
is the $\pi$-phase-kink state (non-Ohmic with nontrivial symmetry 
breaking along the $c$ axis). \cite{Lin2} These proposals share 
the basic framework, \cite{Koyama,Sakai} and difference in 
results is due to choice of the boundary condition. In the present 
article we start from a simplified version \cite{Lin1} of the dynamical 
boundary condition \cite{Bulaevskii} used in Ref.\ 7, 
and gradually introduce effect of the surface impedance. \cite{Lin2}
%
%Emission from intrinsic Josephson junctions (IJJs) such as 
%Bi$_{2}$Sr$_{2}$CaCu$_{2}$O$_{8}$ (BSCCO) has been intensively studied 
%as a candidate of stable source of continuous terahertz electromagnetic wave. 
%Although emission controlled by flow of Josephson vortices \cite{Koyama} 
%had been investigated numerically, \cite{Machida,Tachiki,Lin1} 
%experimental realization of such emission \cite{THz-exp} had been quite 
%difficult. Recently more evident emission from IJJs without external fields 
%was reported experimentally, \cite{Ozyuzer} and two types of emission 
%states were proposed theoretically. One is the McCumber-like state 
%(Ohmic and translational invariant), \cite{Matsumoto} and another 
%is the $\pi$-phase-kink state (non-Ohmic with nontrivial symmetry 
%breaking along the $c$ axis). \cite{Lin2} These proposals share 
%the basic framework, \cite{Koyama,Sakai} and difference in 
%results is due to choice of the boundary condition. In the present 
%article we start from a simplified version \cite{Lin1} of the dynamical 
%boundary condition \cite{Bulaevskii} used in Ref.\ 7, 
%and gradually introduce effect of the surface impedance. \cite{Lin2}

{\it Model and formulation.}
Differently from emission in external fields, dimensional reduction 
along the field cannot be justified anymore in zero-field emission. 
Actually, the above two research groups have already generalized 
their results to three dimensions. \cite{Matsumoto-LT,Hu-Lin} 
Nevertheless, such generalization did not resolve the discrepancy 
between them. Then, in order to clarify the origin of this discrepancy, 
we concentrate on the two-dimensional modeling assuming uniform 
solutions along the $y$ axis. \cite{Matsumoto,Lin2} That is, we solve 
the following differential equations, \cite{Tachiki}
\begin{eqnarray}
\label{eq-1}
\partial_{x'}^{2}\psi_{l+1,l}
&=&(1-\zeta\Delta^{(2)})\left(\partial_{t'}E'_{l+1,l}+\beta E'_{l+1,l}\right.\nonumber\\*
&  &\hspace{1.9cm}\left.+\sin \psi_{l+1,l}-J'\right),\\
\partial_{t'}\psi_{l+1,l}
&=&(1-\alpha\Delta^{(2)})E'_{l+1,l},
\label{eq-2}
\end{eqnarray}
where  the subscript ``$l+1,l$" denotes quantities in the insulating 
layer between the $l$-th and $(l+1)$-th superconducting layers, 
and the operator $\Delta^{(2)}$ is defined in 
$\Delta^{(2)}X_{l+1,l} \equiv X_{l+2,l+1}-2X_{l+1,l}+X_{l,l-1}$. 
The electric field and gauge-invariant phase difference controlled by the dc 
bias current $J$ are basic quantities, and the magnetic field is obtained from 
$\partial_{x'} \psi_{l+1,l}=(1-\zeta\Delta^{(2)})B'_{l+1,l}$.
In these formulas the following scaled quantities are used,
\begin{eqnarray}
\label{red1}
&
x'=x/\lambda_{c},\ t'=\omega_{\rm p}t,\ J'=J/J_{\rm c},
&\\
&
E'_{l+1,l}=\left(\sigma_{c}/(\beta J_{\rm c})\right)E_{l+1,l}^{z},\ 
B'=\left(2\pi\lambda_{c}d/\phi_{0}\right)B,
&\\
&
\zeta=\lambda_{ab}^{2}/(sd),\ \alpha=\epsilon'_{\rm c}\mu^{2}/(sd),\ 
\beta=\sqrt{\epsilon'_{\rm c}}\sigma_{\rm c}\lambda_{c}/(\epsilon_{\rm c}c),
&\\
&
\omega_{\rm p}=c/\left(\sqrt{\epsilon'_{\rm c}}\lambda_{c}\right),\ 
J_{\rm c}=\phi_{0}/(2\pi\mu_{0}\lambda_{c}^{2}d),
&
%&{\displaystyle 
%\zeta=\frac{\lambda_{ab}^{2}}{sd},\ \alpha=\frac{\epsilon'_{\rm c}\mu^{2}}{sd},\ 
%\beta=\frac{\sqrt{\epsilon'_{\rm c}}\sigma_{\rm c}\lambda_{c}}{\epsilon_{\rm c}c},
%}&\\
%&{\displaystyle 
%\omega_{\rm p}=\frac{c}{\sqrt{\epsilon'_{\rm c}}\lambda_{c}},\ 
%J_{\rm c}=\frac{\phi_{0}}{2\pi\mu_{0}\lambda_{c}^{2}d},
%}&
\end{eqnarray}
with the penetration depths $\lambda_{ab}=0.4\mu$m and 
$\lambda_{c}=200\mu$m, thickness of superconducting and 
insulating layers $s=3$\AA\ and $d=12$\AA, respectively, 
Debye length $\mu=0.6$\AA, dielectric constant of the junction 
$\epsilon'_{\rm c}=\epsilon_{\rm c}/\epsilon_{0}=10$ with permittivity 
of the junction $\epsilon_{\rm c}$, plasma frequency $\omega_{\rm p}$, 
conductivity $\sigma_{\rm c}$, flux quantum $\phi_{0}$, and critical current 
$J_{\rm c}$, following the material parameters of BSCCO in Ref.\ 3. 
They give $\alpha=0.1$ and $\beta=0.02$ is taken here.

Width of the junction $L_{x}=86\mu$m is comparable to those of samples used in 
experiments. \cite{Ozyuzer} Since direct simulation of several hundreds of layers 
is difficult, the periodic boundary condition along the $c$ axis corresponding to 
effectively infinite layers is introduced instead. In addition to calculations for the 
number of layers $N=4$, some systems with $N=8$ or $12$ are analyzed to confirm 
numerical consistency. In place of considering outside of IJJs, the dynamical 
boundary condition on edges \cite{Bulaevskii} is introduced. For infinite and uniform 
layers, this boundary condition is simplified \cite{Lin1} as the relation between the 
dynamical part of scaled boundary fields $\tilde{B}'_{l+1,l}$ and $\tilde{E}'_{l+1,l}$:
\begin{eqnarray}
\partial_{x'}\psi_{l+1,l}&=&B'_{\rm ext}+\tilde{B}'_{l+1,l},\\
\partial_{t'}\psi_{l+1,l}&=&\langle E'_{l+1,l}\rangle+\tilde{E}'_{l+1,l},\\
\tilde{E}'_{l+1,l}&=&\mp Z \tilde{B}'_{l+1,l},\ Z=z\sqrt{\epsilon'_{\rm c}/\epsilon'_{\rm d}},
\label{constZ}
\end{eqnarray}
with the dielectric constant of dielectrics $\epsilon'_{\rm d}$. The factor $z$($\geq 1$) 
is considered \cite{Koshelev08a} to appear when the wavelength $\lambda$ of 
emitted electromagnetic wave is much longer than the thickness of junctions 
$L_{z}$, which usually holds in experiments, as $z \approx \lambda/L_{z}$. 
Effect of impedance mismatch on the edges is also included in $Z$. When 
surface fields are not uniform along the $c$ axis, $Z$ depends on wave number 
and frequency in the exact expression. \cite{Bulaevskii,Koshelev08a,Bulaevskii07} 
Width of the sample $L_{x}$ is divided into $80$ numerical grids.

In numerical evaluation of stationary states, procedure for 
parameter sweep is essential. One possible approach is to start 
from a zero-current state and gradually vary currents \cite{Lin2} 
similarly to experiments. 
%%Such a procedure was used in Ref.\ 8.
However, it may not be suitable for study on 
%%investigation of 
stationary states. Simulated time scale is much shorter than that 
in experiments, and accidental trap by metastable condition may 
be held during gradual change of current. Then, 
%%in the present study 
we start from random configurations for each current  and check 
the consistency of results. Continuous data obtained from 
independent initial conditions strongly suggest that the results are 
stationary ones. Since each data point is independent, convergence 
of calculations can be easily checked by additional simulations. 
In order to obtain convergent results from random initial configurations 
precise algorithm with automatically optimized time steps is essential, 
and the RADAU5 ODE solver \cite{radau5} is utilized for this purpose. 

Even if this precise algorithm is used, self-consistent evaluation 
of $\langle E'_{l+1,l}\rangle$ is still difficult. We first fix the static 
value of the surface electric field in order to obtain a stationary state 
under this constraint, and then determine  $\langle E'_{l+1,l}\rangle$ 
self-consistently from this stationary state. When the determined 
average value and amplitude of the surface electric field are denoted 
as $E_{\rm av}$ and $\Delta E$, respectively, strong emission state 
can be obtained from an initial static value $E_{\rm ini}$ satisfying 
$E_{\rm ini}<E_{\rm av}-\Delta E$. This condition looks consistent 
with the fact that strong emission state can only be observed in 
current-increasing process during gradual variance of current. \cite{Lin2}
\begin{figure}
\includegraphics[height=5.5cm]{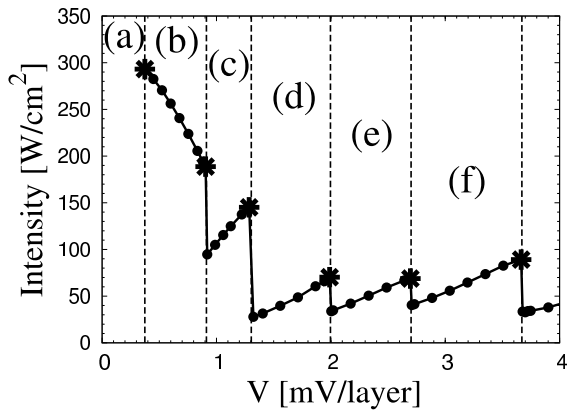}
\caption{\label{fig1}Voltage dependence of emission intensity for $Z=1$. 
Regions (a)$\sim$(f) are divided by jumps of intensity (dashed lines), 
and the data points with star symbols correspond to the voltages at 
which Figs.\ 2 and 3 are drawn.}
\end{figure}
\begin{figure}
\includegraphics[height=6.5cm]{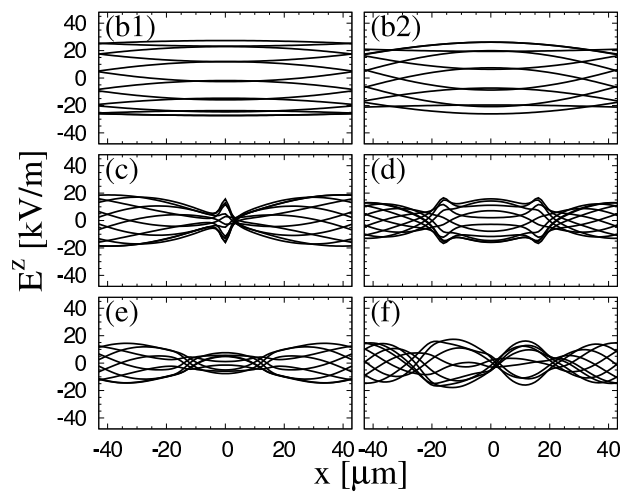}
\caption{\label{fig2}A series of snapshots of the dynamical part 
of electric fields in a layer of IJJ for some typical values of voltages 
(per layer) for $Z=1$: (b1) $0.37$mV, (b2) $0.90$mV, (c) $1.29$mV, 
(d) $1.99$mV, (e) $2.70$mV, and (f) $3.66$mV.}
\end{figure}
\begin{figure}
\includegraphics[height=5.5cm]{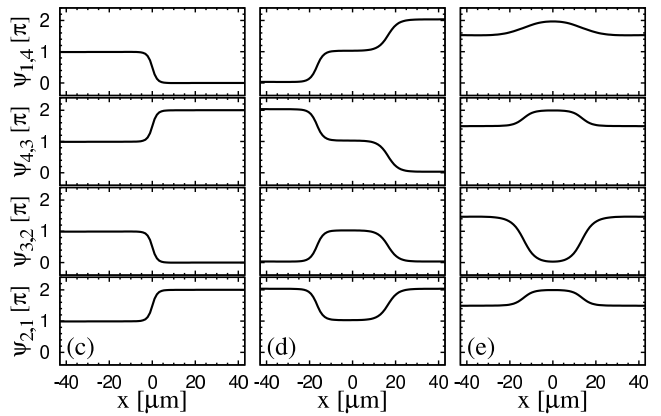}
\caption{\label{fig3}Gauge-invariant phase differences in all insurating 
layers corresponding to the voltages for Figs.\ 2(c), 2(d) and 2(e). 
$\pi/2$- and $3\pi/2$-phase kinks are stabilized in Fig.\ 3(e).}
\end{figure}

{\it Numerical results for $Z=1$.} 
This condition is typically realized for $z=1$ and 
$\epsilon'_{\rm c}=\epsilon'_{\rm d}$, namely no impedance mismatch in 
an infinite system. Voltage dependence of intensity, namely the strength 
of Poynting vector, is shown in Fig.\ \ref{fig1}. Intensity takes maximum 
at the onset of emission, and there exist several abrupt jumps which 
correspond to change of modes of electromagnetic waves in IJJs. 
The range of voltage shown in Fig.\ \ref{fig1} is divided into 6 regions 
(a) to (f), and snapshots of electric field in each region are displayed 
in Figs.\ \ref{fig2}(b1) to \ref{fig2}(f), respectively.

The retrapping region (a) is characterized by vanishing voltage 
for finite bias current. In the region (b), spatial dependence of 
electric fields is small and in-phase motion occurs in all layers, 
which resembles the McCumber state. \cite{Kleiner} 
This McCumber-like state was automatically chosen when translational 
invariance along the $c$ axis is assumed. \cite{Matsumoto} A series of 
snapshots of electric fields for the lowest and highest voltages in this 
region are shown in Figs.\ \ref{fig2}(b1) and \ref{fig2}(b2), respectively. 
They show that spatial dependence of electric fields increases as the 
voltage increases. In the region (c), the state with a $\pi$-phase kink is 
favored. This state accompanies symmetry breaking of phases along the 
$c$ axis (see Fig.\ \ref{fig3}(c)). Such a state was observed for a large 
and complex value of $Z$ \cite{Lin2} or for spatially inhomogeneous 
$J_{\rm c}$ \cite{Koshelev} previously. In the regions (d) and (f), 
similar symmetry-breaking states with increasing number of $\pi$-phase 
kinks are stable (see Fig.\ \ref{fig3}(d): the $2$-kink case with period $4$).
The region (e) is rather special, where two non-$\pi$-phase kinks are 
stabilized. For $N=4$, three layers have $+\pi/2$- and $-\pi/2$-phase 
kinks and one layer has $+3\pi/2$- and $-3\pi/2$-ones (see Fig.\ 3(e)). 
Structure of non-$\pi$-phase kinks depends on $N$. For example, 
for $N=6$ four layers with $\pm 2\pi/3$-phase kinks and two layers with 
$\mp 4\pi/3$-ones. Note that it is not the purpose of the present study 
to check $N$-dependence of emission states in this region in detail.
\begin{figure}
\includegraphics[height=5.5cm]{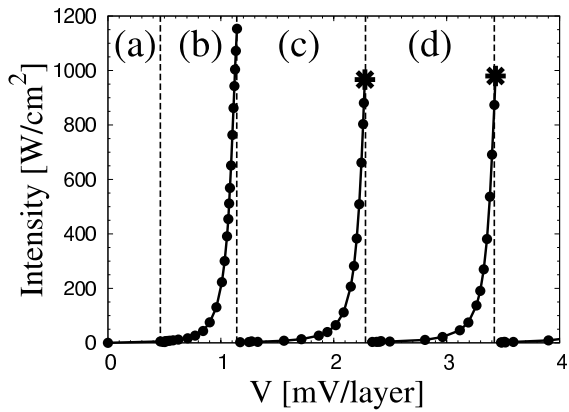}
\caption{\label{fig4}Voltage dependence of emission intensity for 
$Z=10$. Intensity is diverging toward the cavity resonance points 
$V=1.14n$ [mV/layer] (dashed lines) with the integer $n$ to 
specify cavity modes. Other than the retrapping region (a) 
are divided by these voltages, and the data points with star 
symbols correspond to the voltages at which Fig.\ 5 is drawn.}
\end{figure}
\begin{figure}
\includegraphics[height=2.8cm]{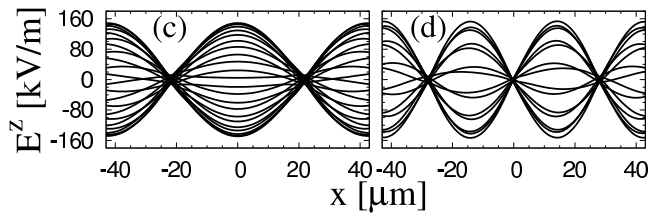}
\caption{\label{fig5}A series of snapshots of the dynamical part 
of electric fields  in a layer of IJJ at intensity peaks for $Z=10$: 
(b) $2.28$mV ($n=2$) and (c) $3.43$mV ($n=3$) 
with the integer $n$ to specify cavity modes.}
\end{figure}
\begin{figure}
\includegraphics[height=5.5cm]{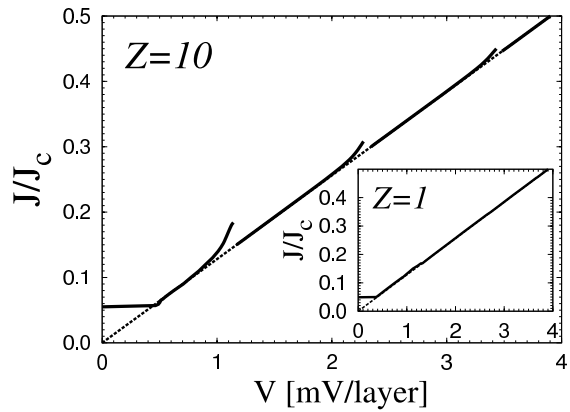}
\caption{\label{fig6}$J$-$V$ curve for $Z=1$ (inset) and $Z=10$. 
Dashed lines represent the Ohm's law for the normal current.}
\end{figure}

{\it Numerical results for $Z=10$.}
Voltage dependence of intensity shown in Fig.\ \ref{fig4} is 
diverging in the vicinity of the cavity resonance points (dashed lines), 
which is quite different from the behavior for $Z=1$ (see Fig.\ \ref{fig1}). 
The McCumber-like state never appears, and standing-wave-like 
behavior (see Fig.\ \ref{fig5}) occurs near  the cavity resonance points. 
Origin of such behavior can be understood from the $J$-$V$ curve shown 
in Fig.\ \ref{fig6}. For $Z=1$ (shown in the inset), the $J$-$V$ curve (solid line) 
almost coincides with the Ohm's law (dashed line), which means that most of 
input current changes to Joule heat. While for $Z=10$, this curve apparently 
goes away from the Ohm's law in the vicinity of the cavity resonance points, 
where the voltage almost saturates when the current increases and the 
excess energy is emitted as electromagnetic waves. 

\begin{figure}
\includegraphics[height=5.5cm]{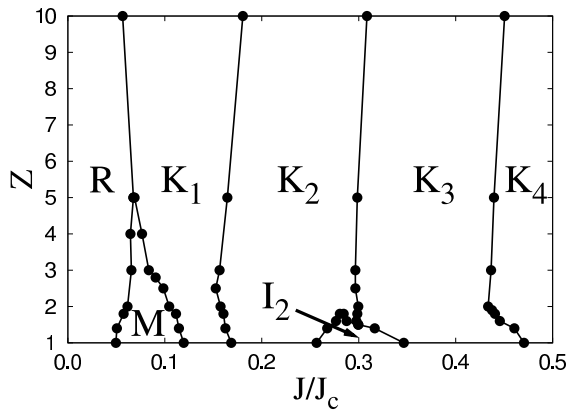}
\caption{\label{fig7}Dynamical phase diagram in the $Z$-$J$ plane. 
R, M, K$_{n}$, and I$_{m}$ denote the retrapping, McCumber-like, 
$n$-$\pi$-phase kink, and  $m$-incommensurate-phase-kink 
states, respectively.}
\end{figure}

{\it Numerical results for other $Z$.}
Emission behaviors for $Z=1$ and $10$ are quite different, 
and it is interesting what happens for intermediate values of $Z$. 
Then, the dynamical phase diagram in the $Z$-$J$ plane 
is given in Fig.\ \ref{fig7}. Apparently, the McCumber-like state 
(denoted by M) is stable only for a limited parameter region: 
$Z<5$ and $0.05<J/J_{\rm c}<0.12$. Boundary of 
the $2$ incommensurate-phase-kink state (denoted by I$_{2}$) 
is evaluated for $N=4$, and quantitative change may occur for 
larger $N$. Important thing is that this phase is stable only for 
$Z<2$, and that most regions of the dynamical phase diagram 
are covered by the $\pi$-phase-kink states (denoted by K$_{n}$, 
$n$: number of kinks). Intensity of emission at the K$_{1}$-K$_{2}$ 
boundary becomes comparable to that at the R-M boundary 
(R: retrapping) for $Z \approx 3$, 
and boundaries between different $\pi$-phase-kink states 
becomes almost independent of $Z$ for $Z\geq 3$, where strong 
emission governed by the ac Josephson relation is observed.

\begin{figure}
\includegraphics[height=5.5cm]{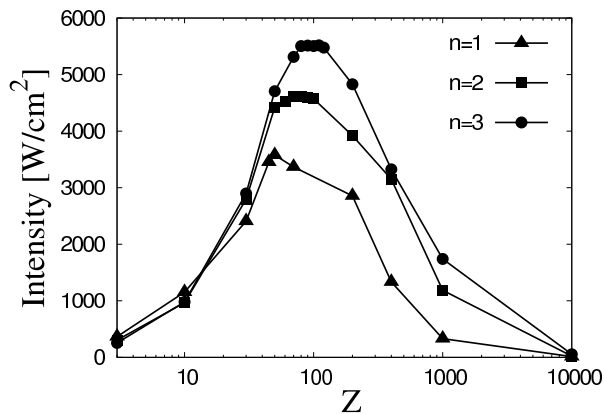}
\caption{\label{fig8}Surface-impedance dependence of maximum 
intensity in a semi-log scale for first $3$ cavity modes: 
$n=1$ (triangles), $2$ (squares), and $3$ (circles).}
\end{figure}

Finally, surface-impedance dependence of emission intensity for larger $Z$ is 
investigated. That is, stationary emission is optimized for various values of $Z$, 
which are varied from $3$ to $10000$. 
Maximum intensity in each cavity mode is observed, and all the maximum values 
are plotted versus $Z$ in Fig.\ \ref{fig8}. The strongest emission is observed 
at $Z\approx 50$ for $n=1$, $Z \approx 80$ for $n=2$, and $Z\approx 100$ 
for $n=3$, respectively. 

{\it Discussions.}
In Ref.\ 8, large surface impedance $|Z|=1000$ was introduced as a 
consequence of large $z$ due to small thickness of the IJJ, \cite{Koshelev08a} 
though this surface effect may be cancelled \cite{Tachiki09} by penetration 
of magnetic fields from transverse directions, which is neglected in 
two-dimensional modeling. Then, stability of standing-wave-like 
behavior for $Z \ge 3$ is important, because emission from an infinite 
IJJ to vacuum ($\epsilon'_{\rm c}=10, \epsilon'_{\rm d}=1$ and $z=1$) 
results in $Z=\sqrt{10}$ and satisfies $Z>3$.

$Z$ and $n$ dependence of emission intensity can be explained as follows: 
Electromagnetic wave in IJJs becomes closer to standing wave for larger $Z$, 
and amplitude of the electric field increases. On the other hand, the amplitude 
cannot exceed its static value and saturates as $Z$ further increases, and 
intensity decreases because the magnetic field is inversely proportional 
to $Z$. As larger value of $n$ is taken, static value of the electric field 
increases and the optimal point of emission shifts toward larger $Z$.

Although phase difference radically varies from layer to layer in the phase-kink 
states as shown in Fig.\ \ref{fig3}, surface fields are almost uniform for any values 
of $Z$. This fact shows that the assumption with constant $Z$ in Eq.\ (\ref{constZ}) 
is a good approximation within the present framework.

{\it Summary.}
In the present article terahertz wave emission from intrinsic Josephson junctions 
without external fields is investigated numerically, and discrepancy between two 
previous studies based on the two-dimensional modeling \cite{Matsumoto,Lin2} 
is resolved. The surface impedance $Z$ is found out to be essential for 
characterization of  emission behavior. For $Z=1$, the McCumber-like state 
is stable for low dc bias currents as reported in Ref.\ 7, and 
this emission state is specific to small $Z$. Even for $Z=1$, $\pi$-phase-kink states 
become stationary for higher currents. Since the $\pi$-phase-kink states require 
symmetry breaking of phases between insulating layers, it was not observed 
in Ref.\ 7, where in-phase motion was assumed {\it a prior}. 
Although large $Z$ is not necessary for stable $\pi$-phase-kink states, 
strong emission from standing-wave-like states does not occur for $Z=1$, 
and such emission is observed at least for $Z \ge 3$. This condition can be 
satisfied only by the impedance mismatch between IJJs and electrodes, 
namely the emission from an infinite IJJ to vacuum gives $Z=\sqrt{10}$. 
Emission behaviors for $Z=10$ are qualitatively similar to those 
in Ref.\ 8 characterized by sharp intensity peaks at the 
cavity resonance points. The strongest emission is observed for 
$Z \approx 50 \sim 100$ slightly depending on the cavity mode $n$, 
and such behavior can be explained by saturation of the electric field.

{\it Acknowledgments.}
The present author would like to thank M.~Tachiki, T.~Koyama, X.~Hu, and S.~Lin for 
helpful comments. This work was partially supported by Grant-in-Aids for Scientific 
Research (C) No.\ 20510121 from JSPS and by the CTC program under JSPS.

\end{document}